\documentclass[11pt,twoside]{article}
\usepackage{./asp2014}

\aspSuppressVolSlug
\resetcounters

\begin{document}

\title{Classical Cepheids in open clusters in the era of Gaia DR2}
\author{Gustavo E. Medina,$^1$ Bertrand Lemasle,$^1$ Eva K. Grebel,$^1$ Steffi X. Yen$^2$}

\affil{$^1$Astronomisches Rechen-Institut, Zentrum f{\"u}r Astronomie der Universit{\"a}t Heidelberg, 69120 Heidelberg, Germany \\
\email{gustavo.medina@uni-heidelberg.de}}
\affil{$^2$Landessternwarte, Zentrum f{\"u}r Astronomie der Universi{\"a}t Heidelberg, 69117 Heidelberg, Germany \\
}

\paperauthor{Gustavo E. Medina}{gustavo.medina@uni-heidelberg.de}{0000-0003-0105-9576}{Zentrum f{\"u}r Astronomie der Universit{\"a}t Heidelberg}{Astronomisches Rechen-Institut}{Heidelberg}{}{69120}{Germany}

\paperauthor{Bertrand Lemasle}{lemasle@uni-heidelberg.de}{}{Zentrum f{\"u}r Astronomie der Universit{\"a}t Heidelberg}{Astronomisches Rechen-Institut}{Heidelberg}{}{69120}{Germany}

\paperauthor{Eva K. Grebel}{grebel@ari.uni-heidelberg.de}{0000-0002-1891-3794}{Zentrum f{\"u}r Astronomie der Universit{\"a}t Heidelberg}{Astronomisches Rechen-Institut}{Heidelberg}{}{69120}{Germany}

\paperauthor{Steffi X. Yen}{s.yen@lsw.uni-heidelberg.de}{0000-0001-8405-488X}{Zentrum f{\"u}r Astronomie der Universit{\"a}t Heidelberg}{Landessternwarte}{Heidelberg}{}{69117}{Germany}

\begin{abstract}
Classical Cepheids in open clusters are key ingredients for stellar population studies and the characterization of variable stars, as they are tracers of young and massive populations and of recent star formation episodes. 
Cluster Cepheids are of particular importance since they can be age dated by using the cluster's stellar population to obtain the Cepheid period-luminosity-age relation.
In this contribution, we present the preliminary results of an all-sky search for classical Cepheids in Galactic open clusters by taking advantage of the unprecedented astrometric precision of the second data release of the Gaia satellite. 
To do this, we determined membership probabilities by performing a Bayesian analysis based on the spatial distribution of Cepheids and clusters, and their kinematics.
Here we describe our adopted methodology.
\end{abstract}

\section{Introduction}
The search for Cepheids in coeval stellar systems formed under the same physical conditions, such as open clusters, has been the subject of numerous studies over the last decades \citep[e.g.][]{Anderson13}, owing to their status as valuable calibrators of fundamental relations of these variables. 
In this work, we study classical Cepheids that are members of open clusters. 
We aim to quantify the membership probability of these Cepheids by using updated kinematic information (of both clusters and Cepheids) based on the unparalleled data from the Gaia data release 2 \citep[DR2; ][]{Gaia18}, and recent radial velocity studies. 
Our final goal is to use these coeval ensembles to independently determine the age of the Cepheids they host, and use this information to constrain their period-luminosity-age relation and its dependence on metallicity.

\section{Methodology and Results}
We reexamine the membership of classical Cepheids in Galactic open clusters, based mostly on the Gaia DR2 Cepheids (using the reclassifications made by \citealt{Ripepi19}, which comprise $\sim$ 800 stars) and open cluster databases such as the catalogs of \citet{Dias02} and \citet{K13}. 
By combining these databases with more recent studies \citep{Ferreira19, CG19, Torrealba19}, we built a catalog containing a total of $\sim$ 3500 clusters for our analysis.

By following a Bayesian approach, membership probabilities are computed using updated kinematics of both Cepheids and clusters (proper motions, parallaxes, and radial velocities), and their projected on-sky distances 
\citep[e.g.][]{CG18a, CG18b,Carrera19,Melnik15}. 
To determine the posterior probabilities $P(A|B)$, we chose a prior $P(A)$ that falls exponentially with the on-sky separation between a Cepheid and its corresponding cluster.
For the likelihood we adopted the Mahalanobis distance, which can be used as a measure of the multidimensional distance, in this case based on kinematics, between a point (a Cepheid) and a distribution (an open cluster).
More details of the usage of the Mahalanobis distance in this context can be found in the work by \citet{Anderson13}. 

Applying this method to our sample of clusters and Cepheids, we found $\sim$ 60 possible associations, including previously known Cepheid-cluster pairs and new candidates, such as the cluster Gaia 5 and the Cepheid V0423 CMa.
Table 1 shows an extract of our results.
Using PARSEC isochrones \citep{Bressan12}, we also fitted evolutionary models to the color-magnitude diagrams of a selection of clusters to determine their ages, metallicities, distances and reddening values. The next steps of this project involve the analysis of the cluster parameters and Cepheid properties to constrain the period-luminosity-age relation of these variables.

\begin{table}[!ht]
\caption{
Extract of the cluster – Cepheid pairs with highest membership probabilities.}
\smallskip
\begin{center}
{\small
\begin{tabular}{cccccc}  
\tableline
\noalign{\smallskip}
Open cluster & Cepheid & Constraints & P(A) & P(B|A) & P(A|B):\\
\noalign{\smallskip}
\tableline
\noalign{\smallskip}
NGC 6087 & S Nor & $\varpi$, RV, $\mu_{\alpha}^{*}$, $\mu_{\delta}$ & 1.0 & 0.99 & 0.99 \\
NGC 6067 & V0340 Nor & $\varpi$, $\mu_{\alpha}^{*}$, $\mu_{\delta}$ & 1.0 & 0.98 & 0.98 \\
IC  4725 & U Sgr & $\varpi$, RV, $\mu_{\alpha}^{*}$, $\mu_{\delta}$ & 1.0 & 0.75 & 0.75 \\
BH 72 & DP Vel & $\varpi$, $\mu_{\alpha}^{*}$, $\mu_{\delta}$ & 0.32 & 0.66 & 0.21 \\
King 4 & UY Per & $\varpi$, $\mu_{\alpha}^{*}$, $\mu_{\delta}$ & 0.31 & 0.24 & 0.07 \\
\noalign{\smallskip}
\tableline\
\end{tabular}
}
\end{center}
\end{table}

\acknowledgements 
GM would like to acknowledge the support of the Hector Fellow Academy.


\begin{thebibliography}{}
\bibitem[Anderson et al.(2013)]{Anderson13}
Anderson R. I., Eyer L., Mowlavi N., 2013, MNRAS, 434, 2238. 

\bibitem[Bressan et al.(2012)]{Bressan12} Bressan, A. et al., 2012, MNRAS, 427, 127.
\bibitem[Cantat-Gaudin et al.(2018a)]{CG18a} Cantat-Gaudin T., et al., 2018, A\&A, 615, A49
\bibitem[Cantat-Gaudin et al.(2018b)]{CG18b} Cantat-Gaudin T., et al., 2018, A\&A, 618, A93.
\bibitem[Cantat-Gaudin et al.(2019)]{CG19} Cantat-Gaudin T., et al., 2019, A\&A, 624, A126.
\bibitem[Carrera et al.(2019)]{Carrera19} Carrera R., et al., 2019, A\&A, 623, A80.
\bibitem[Dias et al.(2002)]{Dias02} Dias W. S., et al., 2002, A\&A, 389, 871.
\bibitem[Ferreira et al.(2019)]{Ferreira19} Ferreira F. A., et al., 2019, MNRAS, 483, 5508.
\bibitem[Gaia Collaboration et al.(2018)]{Gaia18} Gaia Collaboration et al., 2018, A\&A, 616, A1.
\bibitem[Kharchenko et al.(2013)]{K13} Kharchenko N. V., et al., 2013, A\&A, 558, A53
\bibitem[Mel'nik et al.(2015)]{Melnik15} Mel'nik A. M., et al., 2015, AN, 336, 70
\bibitem[Ripepi et al.(2019)]{Ripepi19} Ripepi V., et al., 2019, A\&A, 625, A14
\bibitem[Torrealba et al.(2019)]{Torrealba19} Torrealba G., Belokurov V., Koposov S., 2019, MNRAS, 484, 2181

\end{thebibliography}
\end{document}